\documentclass[a4paper,11pt]{article}
\pdfoutput=1 

\usepackage{jinstpub} 

\usepackage{soul}
\usepackage{color}
\usepackage{multicol} 
\usepackage{array}
\usepackage{wasysym}
\usepackage{hyperref}
\usepackage{eurosym}
\usepackage{pifont}
\usepackage{subfigure}
\usepackage[bottom]{footmisc}
\usepackage{lineno}

\title{\boldmath A Light Calibration System for the ProtoDUNE-DP Detector}

\author[a]{D.~Belver,}
\author[b]{J.~Boix,}
\author[a]{E.~Calvo,}
\author[a,1]{C.~Cuesta,,\note{Corresponding author}}
\author[a]{A.~Gallego-Ros,}
\author[a]{I.~Gil-Botella,}
\author[a]{S.~Jim\'{e}nez,}
\author[a]{C.~Lastoria,}
\author[b]{T.~Lux,}
\author[a]{I. Mart\'{i}n,}
\author[a]{J.J. Mart\'{i}nez,}
\author[a]{C.~Palomares,}
\author[a]{D.~Redondo,}
\author[a]{J.~Soto-Oton,}
\author[b]{D.~Vargas,}
\author[a]{A.~Verdugo}

\affiliation[a]{Centro de Investigaciones Energ\'{e}ticas, Medioambientales y Tecnol\'{o}gias (CIEMAT),\\ Madrid, Spain}
\affiliation[b]{Institut de F\'{i}sica d\'{}Altes Energies (IFAE) - Barcelona Institute of Science and Technology (BIST),\\   Bellaterra (Barcelona), Spain}

\emailAdd{clara.cuesta@ciemat.es}

\abstract{A LED-based fiber calibration system for the ProtoDUNE-Dual Phase (DP) photon detection system (PDS) has been designed and validated. ProtoDUNE-DP is  a 6x6x6\,m$^3$ liquid argon time-projection-chamber (LAr TPC) currently being installed at the Neutrino Platform at CERN. The PDS is based on 36 8-inch photomultiplier tubes (PMTs) and will allow triggering on cosmic rays. The system serves as prototype for the PDS of the final DUNE DP far detector in which the PDS also has the function to allow the 3D event reconstruction on non-beam physics. For this purpose an equalized PMT response is desirable to allow using the same threshold definition for all PMT groups, simplifying the determination of the trigger efficiency. The light calibration system (LCS) described in this paper is developed to provide this and to monitor the PMT performance in-situ.}


\begin{document}
\maketitle
\flushbottom

\section{Introduction}
\label{sec:intro}

The DUNE experiment aims to address key questions in neutrino physics and astroparticle physics~\cite{duneCDRv2,duneIDRv1}. It includes precision measurements of the parameters that govern neutrino oscillations with the goal of measuring the CP violating phase and the neutrino mass hierarchy with a muon neutrino beam produced at Fermilab. The physics program also addresses non-beam physics as nucleon decay searches and the detection and measurement of the electron neutrino flux from a core-collapse supernova within our galaxy. DUNE will consist of a near detector placed at Fermilab close to the production point of the muon neutrino beam of the Long-Baseline Neutrino Facility (LBNF), and four 10\,kt fiducial mass LAr TPCs as far detector in the Sanford Underground Research Facility (SURF) at 4300\,m.w.e. depth at 1300\,km from Fermilab~\cite{duneCDRv4,duneIDRv2,duneIDRv3}. 

In order to gain experience in building and operating such large-scale LAr detectors, an R\&D program is currently underway at the CERN Neutrino Platform~\cite{ProtoDUNEs}. It consists of two prototypes with the specific aim of validating the design, assembly, and installation procedures, the detector operations, as well as data acquisition, storage, processing, and analysis. The two prototypes employ LAr TPCs as detection technology. One prototype only uses LAr, called ProtoDUNE Single-Phase (SP)~\cite{ProtoDUNESPtdr}, and the other uses argon in both its gaseous and liquid state, thus the name ProtoDUNE-DP~\cite{wa105}. Both detectors have similar sizes. 

ProtoDUNE-DP has an active volume of 6$\times$6$\times$6 m$^{3}$ corresponding to a fiducial mass of 300\,t. In ProtoDUNE-DP the charge is extracted, amplified, and detected in gaseous argon above the liquid surface allowing a finer readout pitch, a lower energy threshold, and better pattern reconstruction of the events. In addition, the scintillation light signal is used as trigger for non-beam events, to determine precisely the event time, needed for a full 3D reconstruction of non-beam events,  and for cosmic background rejection. There might be also a possibility to perform particle identification exploiting the detected light signals. The PDS of ProtoDUNE-DP~\cite{protoDUNElight} is formed by 36 8-inch cryogenic PMTs (R5912-02MOD from Hamamatsu) placed below the cathode grid. As wavelength-shifter, tetraphenyl butadiene (TPB) is coated directly on the PMTs. The ProtoDUNE-DP PMTs have been validated and characterized~\cite{protoDUNEPMTs}.

This paper describes the ProtoDUNE-DP LCS design and validation. One of the main goals of the PDS is to provide triggering for non-beam physics. The trigger is based on the amplitude of PMT-signals. The amplitudes of the PMT-signals are summed for groups of certain PMTs and/or for all PMTs and then, these input signals are discriminated according to the trigger logic. An equalized PMT response allows to use the same threshold definition for all PMT groups, simplifying the determination of the trigger efficiency. Beside measuring the PMT gain, it is also designed to monitor the stability of the PMT response, and so its quantum efficiency, during the performance of ProtoDUNE-DP. As concluded from the operation of a previous 3$\times$1$\times$1\,m$^3$ LAr TPC detector~\cite{311} and the ProtoDUNE-DP PMT characterization~\cite{protoDUNEPMTs}, a LCS is strongly recommended during the data taking period. The conceptual design is detailed in Sec.~\ref{sec2}. The characterization measurements performed individually for the different components are presented in Sec.~\ref{sec3}. Sec.~\ref{sec4} describes the results of the complete system validation. Finally, the projection of this system to the DUNE DP far detector~\cite{duneIDRv3} is explained in Sec.~\ref{sec6}.

\section{Design description}
\label{sec2}

The design of the LCS must take into account that it will be operated at cryogenic temperature (CT) ($\sim$87\,K) and must include the minimal number of elements inside the cryostat, so the light source is operated from outside with the constrain of a small number of available feedthroughs. The LCS must illuminate 36 PMTs at Single Photo-Electron (SPE) level and at larger light intensities to ensure that the gain calibration and the linearity can be determined from the data. The acquisition rate is not a strong requirement since the rate should be kept within $\sim$kHz to avoid the PMT fatigue~\cite{protoDUNEPMTs}.

\begin{figure}[ht]
\bigskip
\centering
 \includegraphics[height=0.5\textwidth]{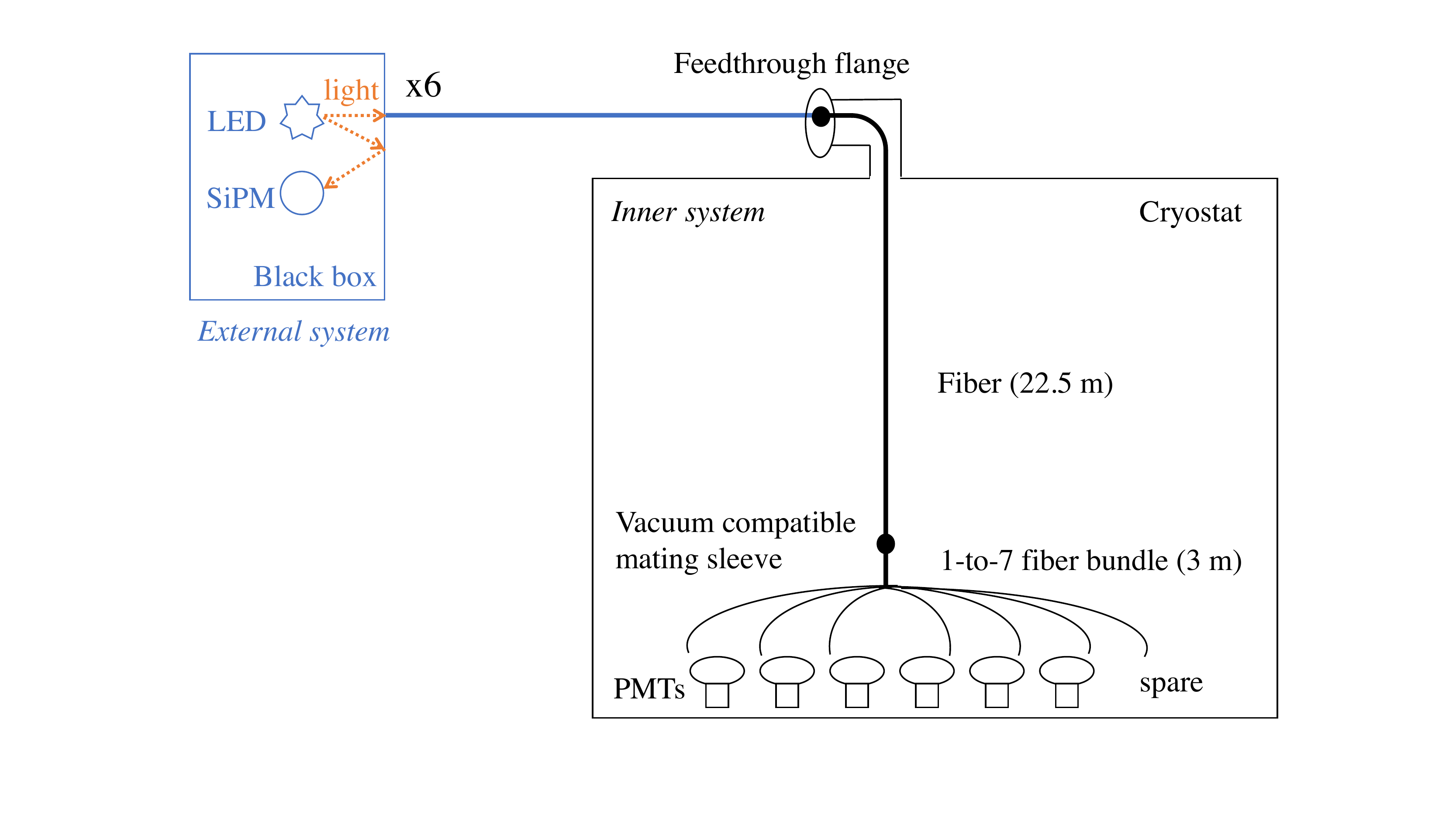}
\caption{Sketch of the ProtoDUNE-DP LCS. The external system at RT is shown in blue and the inner system at CT in black.}
\label{LCS}
\end{figure}

A LED-driven fiber calibration system~\cite{LCSuboone,LCSborexino,LCSminos} is designed so that a configurable amount of light reaches each PMT. The calibration light is provided by a blue LED of 465\,nm (matching the PMT maximum quantum efficiency wavelength) using a Kapustinsky~\cite{1985NIMPA.241..612K} circuit as LED driver, and transmitted by a fiber system ending with an optical SMA connector installed on each PMT support structure (see Fig.~\ref{LCS}). There are six LEDs placed in a hexagonal geometry and a reference sensor at the center to check the LEDs performance. The direct light goes to the fiber, and the stray light to the SiPM used as reference sensor. Each LED is connected to an external fiber going to one feedthrough. Then, six fibers are connected inside the cryostat and each one of these fibers is attached to a 1-to-7 fiber bundle, so that one fiber is finally installed pointing at each PMT. The PMTs are oriented with the first dynode perpendicular to the Earth magnetic field and the fiber parallel to the first dynode to avoid the magnetic field moves the photo-electron away from the first dynode reducing the PMT gain \cite{Calvo:2009wn}. In the following, the components placed outside the cryostat at room temperature (RT) form the \textit{external system} and the ones installed inside it at cryogenic temperature (CT) the \textit{inner system}. 

\subsection{External system}
The design of the external components is driven by the aim to have a cost-effective system which could serve as prototype for the final LCS of the DUNE DP far detector. The goal is to get a light source of adjustable and reproducible intensity. As the stability of the LED is not guaranteed, a reference light sensor monitors the amount of the injected light. The light is injected in form of several ns-long pulses provided by a Kapustinsky circuit.

\begin{figure}[h!]
\centering
\subfigure[]{\includegraphics[width=0.25\textwidth]{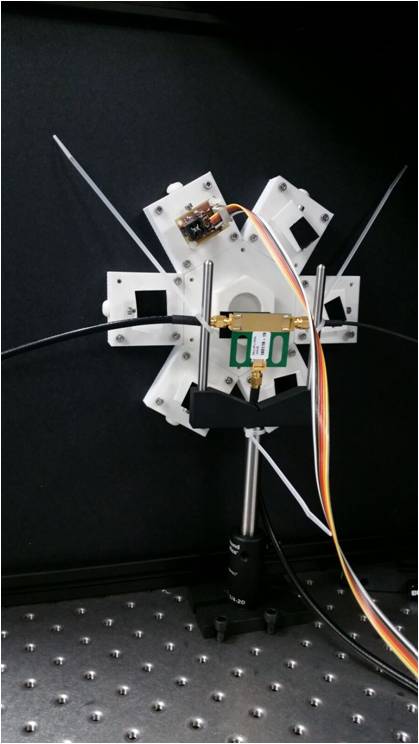}\label{fig:lightguide}}
\subfigure[]{\includegraphics[width=0.55\textwidth]{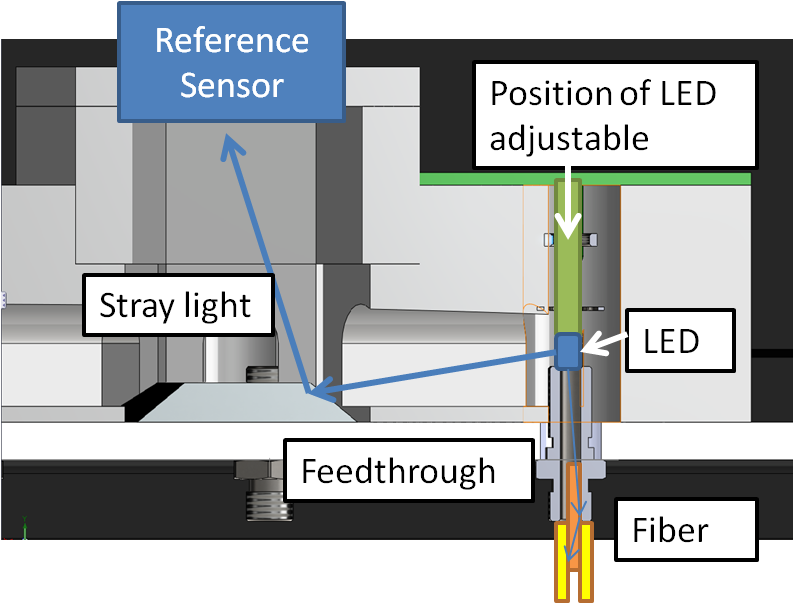}\label{fig:stray}}
\caption{(a) Photo of the light guide structure during the development phase. The 6 arms are visible, and on one of them a prototype LED driver is mounted. During these tests, the SiPM was mounted on a commercial development board. (b) Schematics of the light path from the LED to the reference sensor.}
\label{fig_source}
\end{figure}

The setup consists of a commercial black box, XE25C9D/M from Thorlabs\footnote{www.thorlabs.com}, in which a custom light guide structure is used for the light distribution among the different components (see Fig.~\ref{fig_source}). The light guide structure was produced by means of 3D printing. It has a central part and 6 arms. The reference sensor is located on the central part, and, on each of the arms a PCB with the Kapustinsky circuit is mounted. The circuit is identical to the original design besides of a low pass filter to minimize the effect of the electrical pulses in the electronics control. The LED, NSPB300B from Nichia Corp.\footnote{www.nichia.com}, with a peak wavelength of 465\,nm is placed on the PCB in front of an optical SMA feedthrough. On the other side of each feedthrough an optical fiber is connected. It transports the light to one of the 6 feedthroughs in the instrumentation flange on top of the cryostat. To transport the light to the feedthrough, an optical fiber, FG105LCA-CUSTOM-MUC from Thorlabs with a length of 9.5\,m and fiber diameter of 105\,$\mu$m is chosen due to the optimal performance during the full system validation (see Sec. \ref{sec4}). While a large fraction of the LED light is emitted in the forward direction, a small fraction, the stray light, is emitted under a large angle and reaches by reflection on the 3D printed walls the central region of the light guide structure where it is detected by a SiPM, MicroFJ-30035-TSV-TA, from SensL\footnote{www.sensl.com}. 

The SiPM sensor provides two signal outputs: a fast output and a slow output. The fast output, connected directly to an electrical SMA feedthrough, provides a signal of a few ns width while the slow output provides a signal with a few ns rise time and a long tail of some hundred ns. The slow output signal is used in the onboard integrator circuit and also goes to a SMA connector and can be used by an external digitizer. The SiPM sensor is operated with a bias voltage of about 30\,V. 

The SiPM sensor and the signal shaper/integrator are assembled on a PCB that fits in the central hole of the light guide structure. The PCB is in charge of distributing the trigger and the LED bias voltage coming from the control electronics to the Kapustinsky circuit through flat cables. This board is connected through a DB25 connector to a BeagleBone Black module\footnote{https://beagleboard.org/black}.

The BeagleBone Black module is mounted on the outside of the black box and contains the control software and firmware. This mini-PC provides the voltage pulse of 3.3\,V which triggers the light pulse and controls the low voltage (LV) power supply, CPX200DP, from Atti\footnote{www.aimtti.com}, delivering the DC voltage of the LED driver and the SiPM bias voltage. In addition, its built-in 12-bit ADC digitizes and calculates the integral of the light pulse measured by the SiPM. The operation parameters,  as the pulses rate and the LED bias voltage, can be changed via Ethernet connection. The signal integral values are read via this Ethernet connection.

To distribute and concentrate the signals, a shield board has been developed. This board is in charge of collecting signals form the Beaglebone and concentrate them in a DB25 connector. It also contains an inverter buffer for flexibility reasons allowing the usage of different ADCs for the readout. The board also includes the trigger distribution circuit.

\subsection{Inner system}
\begin{figure}[h!]
\centering
\includegraphics[height=0.35\textwidth]{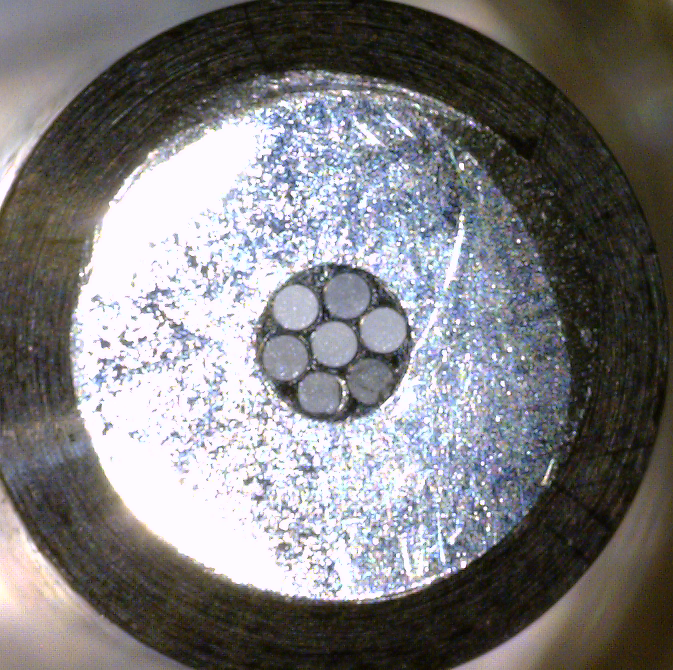}
\caption{1-to-7 fiber bundle end where the 7 fibers can be seen. \label{fig_fibers_b} }
\label{fig_fibers}
\end{figure}

The inner system is designed to maximize the light transmission of the PMTs at CT. The external fibers are connected to 6 female optical feedthroughs from Allectra\footnote{www.allectra.com} installed at 2 flanges. Inside the cryostat, a single 22.5\,m-long fiber
goes down from each optical feedthrough routed along the walls of the cryostat to its bottom where a 1-to-7 fiber bundle (see Fig.~\ref{fig_fibers_b}) is connected to each long fiber. In total, 36 of these fibers are guided to the PMTs at the bottom of the detector. The end of the fiber is fixed at the PMT support structure pointing to the photocathode. Details of the components can be found in Table~\ref{tab_components}. 

\begin{table}[h]
\centering
\begin{tabular}{|c|c|c|c|c|c|c|c}
\hline
Qty.  & Component & Manufacturer & Reference & \diameter	& Length & Jacket \\
	 & 			 & 			  	& 			& ($\mu$m)	& (m) 	 &  \\
\hline
\hline
2 & Feedthrough flange & Allectra  & 152-FCF-C40-3 & 3200 & - & -\\
\hline
6 &Inner Fiber & Thorlabs &FT800UMT& 800 & 22.5 & SS\\
\hline
6 & Mating sleeve & Thorlabs & ADASMAV & 3200 & - & -\\ 
\hline6 & 1-to-7 fiber bundle & Thorlabs & FT200UMT & 200 each & 3 & SS/black$^1$\\
\hline
\end{tabular}
\caption{Description of the fibers, feedthroughs and connectors used in the LCS inner system. $^1$Stainless-steel (SS) jacket are used at the common end (25\,cm) and black jacket at the split ends.}
\label{tab_components}
\end{table}

The fibers and bundles are 0.39\,NA (numerical aperture) TECS$^\text{TM}$ hard-clad, multimode, step-index fibers with high OH (hydroxyl) ion concentration to increase the light transmission at low wavelengths. In order to optimize the light transmission of the fiber-bundle connection, the inner fibers have a diameter of 800\,$\mu$m, big enough to distribute uniformly the light at the bundle entrance (total diameter 700\,$\mu$m) given the NA. From the mechanical point of view, the described approach of bundles attached to fibers is safer than connecting directly the bundles to the feedthroughs. However, as connections may produce certain light loss, a dedicated characterization of the components is carried out to validate this option, see Sec.~\ref{sec3.2}. The bundle length is minimized (3\,m) and the length of the fiber (22.5\,m) is determined considering the cryostat dimensions and the minimum length required to route the fibers through the detector. 

In order to have a good light homogeneity at the fiber-bundle, SMA connectors are chosen. The feedthroughs are high vacuum rated, but we performed vacuum and pressure tests to verify the flange tightness even though ProtoDUNE-DP will not operate in vacuum conditions. Vacuum compatible SMA to SMA mating sleeves (MS) are required to avoid that water vapor freezes inside the connector which would reduce the light transmission.

\section{Component characterization}
\label{sec3}
A systematic characterization of the external and internal components is carried out in the laboratory to evaluate the feasibility of the system for ProtoDUNE-DP.

\subsection{External system characterization}
\label{sec3.1}
The external components, the LED driver PCBs and the central PCB with the Sensl SiPM are characterized using a calibrated photodiode\footnote{818-UV/DB from Newport Corp.} and a 1\,inch PMT (model: 9116 A)~\cite{958753}. The photodiode provides the average power over many light pulses while the PMT signal is used to study the pulse shape of the different LED driver PCBs. 

\subsubsection{LED frequency scan}
\begin{figure}[h]
\bigskip
\centering
\subfigure[]{\includegraphics[width=0.45\textwidth]{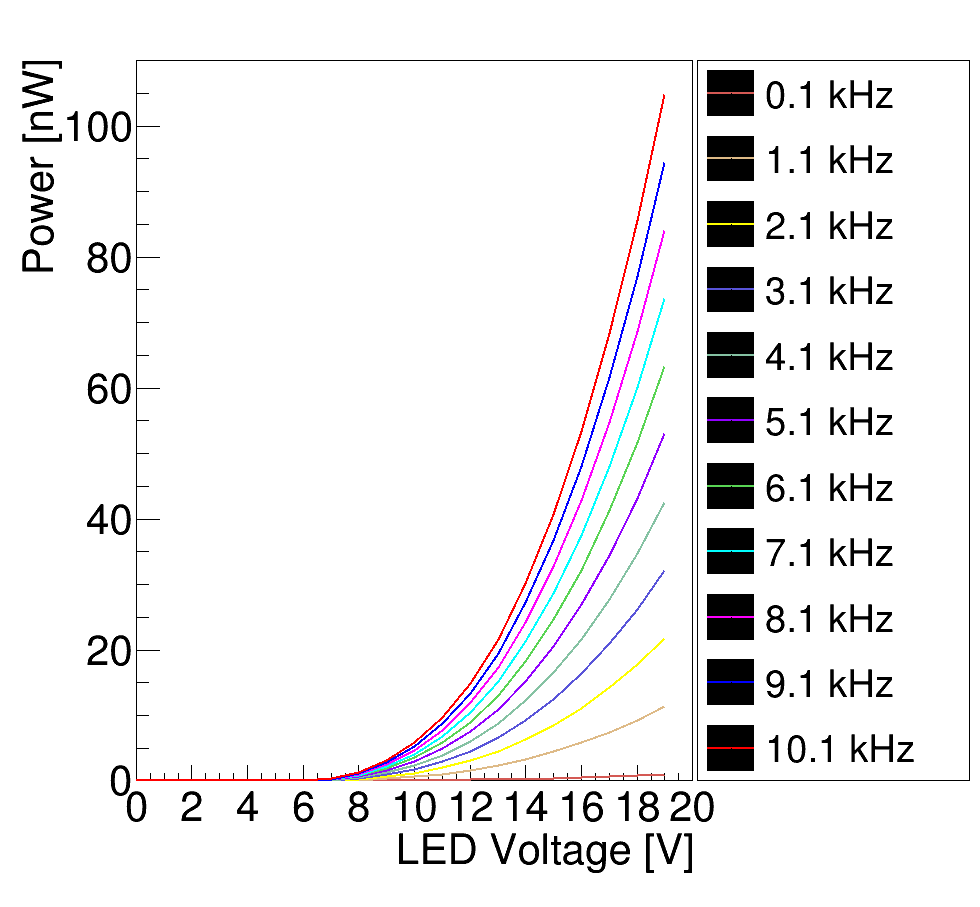}\label{fig:I_Freq}}
\subfigure[]{\includegraphics[width=0.45\textwidth]{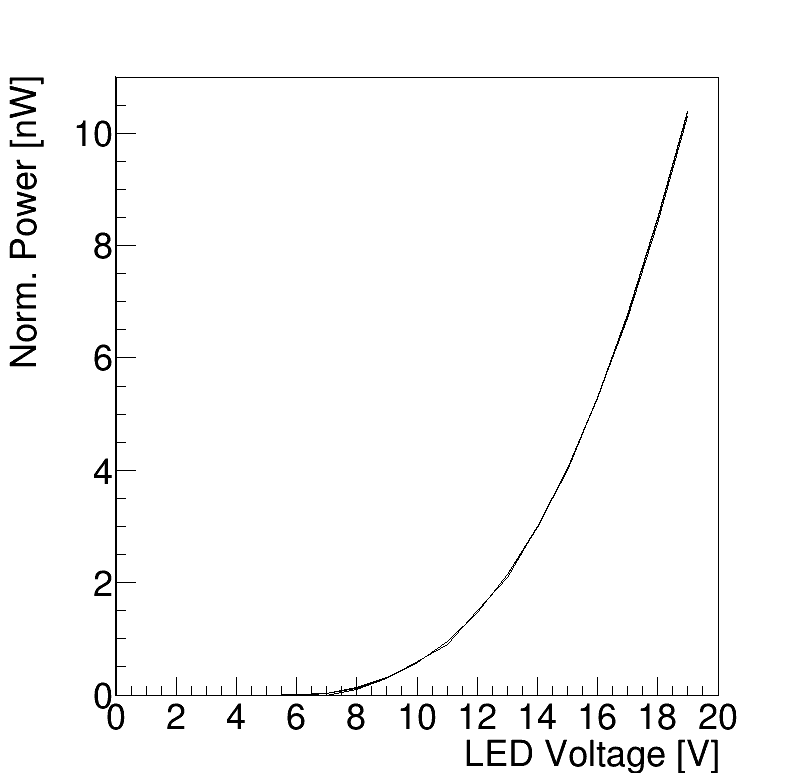}\label{fig:I_Freq_norm}}
\caption{(a) Output power as a function of the voltage applied to the LED for different frequencies between 100 and 10100 Hz in 1000 Hz steps. (b) Same curves normalized to 1 kHz showing that there is not saturation effect in this frequency range. }
\end{figure}

It is studied if the repetition rate of the light pulses leads to any saturation effect. For this study the LED drivers are triggered with input pulses between 100\,Hz and 10100\,Hz repetition rate in 1\,kHz steps. The optical fiber is directly connected to the powermeter using a suitable SMA adapter. The output curves are recorded for LED voltages between 0 and 19.5 V. The curves for one of the LED drivers are shown in Fig.~\ref{fig:I_Freq}. Fig.~\ref{fig:I_Freq_norm} shows the same curves when the output power is normalized to that of 1 kHz. As one can see, all curves lay over each other indicating that the drivers show no frequency depending effect within the tested range. The small difference for one of the curves comes from the 100 Hz measurement where the output power is close to the sensitivity limit of the powermeter. 

\subsubsection{LED output power spread}
The same data are used to study the spread of the output power for the different LED driver PCBs. For an LED voltage of 19\,V, the spread over all PCBs is found to be 30\%. One has to note that this number includes the spread related to the position of the PCB. To investigate the spread related to the positioning of the LED driver in the light guide structure alone, one of the PCBs is installed and removed several times in the same arm, and afterwards, it is installed in the other arms and the output power is measured for all configurations. The measured spread coming from the positioning is $\pm$10\%. However, this output spread has no impact on the performance of the LCS since it does not affect the SPE spectrum. 

\subsubsection{Time stability of the LED driver}
\begin{figure}[h]
\bigskip
\centering
\subfigure[]{ \includegraphics[width=0.45\textwidth]{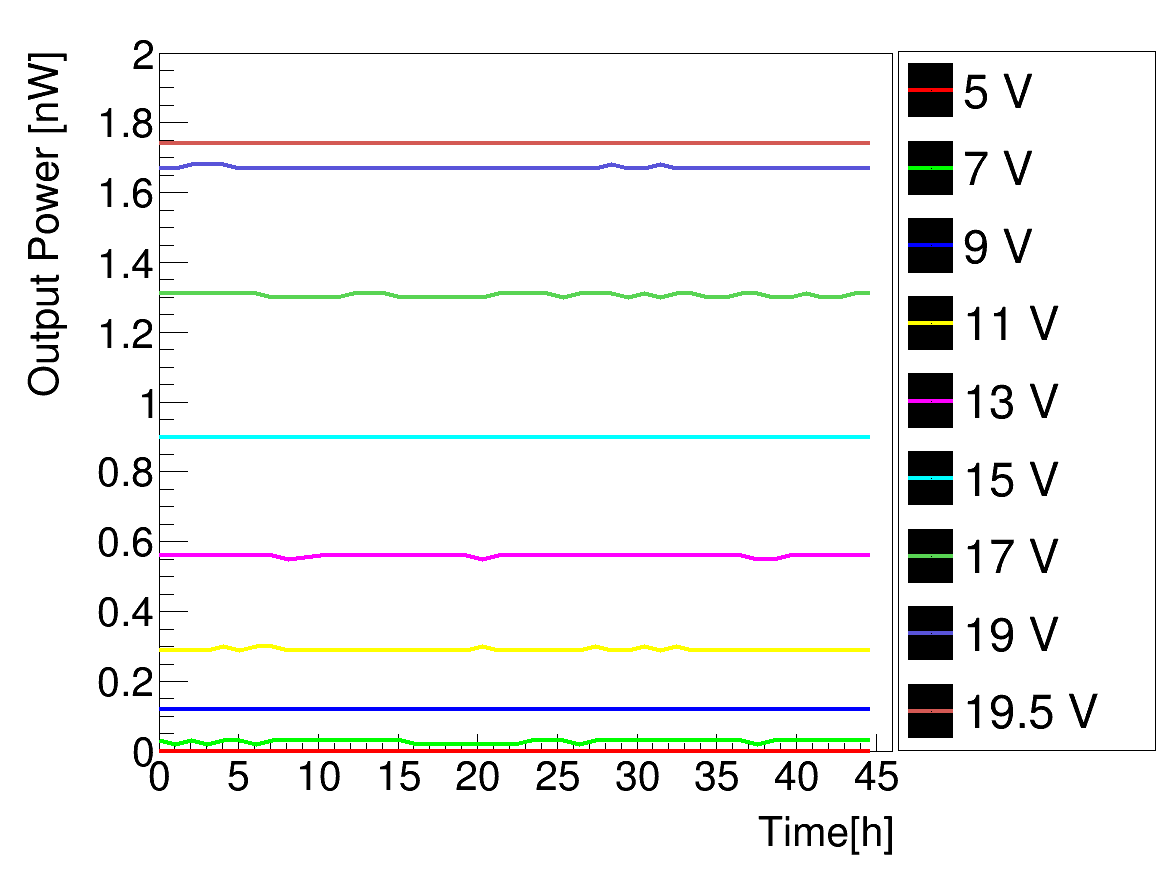}\label{fig:I_Ch_TimeStab}}
\subfigure[]{\includegraphics[width=0.4\textwidth]{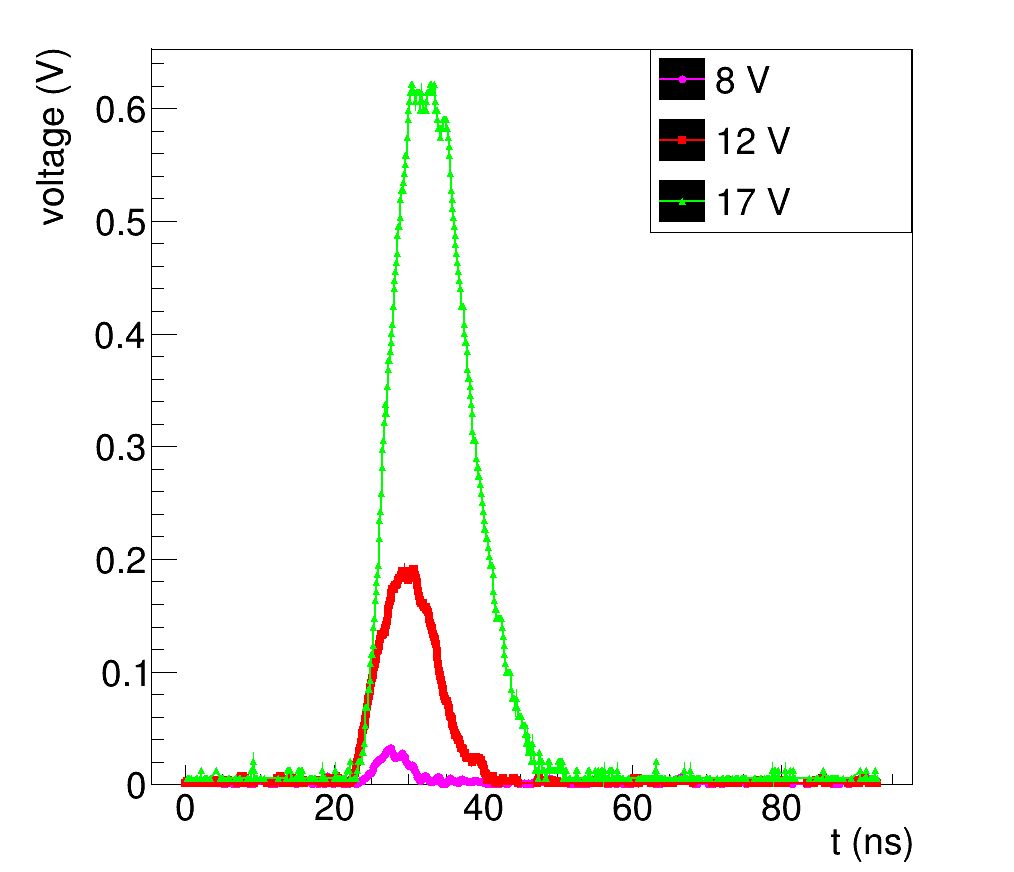}\label{fig:I_Ch_PulseShape}}
\caption{(a) The output power measured for one of the LED drivers for various LED bias voltages at 1\,kHz. The curves correspond from bottom to top to voltages of 5 (red), 7 (green), 9 (blue), 11 (yellow), 13 (magenta), 15 (turquoise), 17 (green), 19 (blue) and 19.5 (black) V. (b) Typical shape of the light pulses for 3 different LED bias voltages (8 (magenta), 12 (red), 17 (green) V). The width of the pulses increases with the LED bias voltage to maximal around 30\,ns.}

\end{figure}

For the study of the time stability of the LED driver, one of the drivers is placed closely in front of the photodiode, and the output power of the LED is measured every hour during almost 45 hours. For each measurement a scan of the LED bias voltage is performed starting from 0\,V and ramping up to 19.5\,V in 0.5\,V steps. The repetition rate for the LED pulses is set to 1\,kHz. The result is shown in Fig.~\ref{fig:I_Ch_TimeStab} for various LED bias voltages. The variation of the output power is found to be stable within the precision of the powermeter of 0.01\,nW, and  for none of the LED bias voltages a systematic drift of the output power over time is observed. The maximal variations of the temperature in the acclimatized room are below $\pm2^\textrm{o}$\,C.


\subsubsection{LED pulse shape}
For the pulse shape study the LED driver and the PMT are mounted in a black box.  To reduce the light yield to an acceptable level for the PMT, a diffuser is placed in front of the LED at about 40\,cm from the PMT. The pulse shape is measured for all LED drivers and various LED bias voltages between 0 and 17\,V. Fig.~\ref{fig:I_Ch_PulseShape} shows the typical pulse shape for 3 different LED bias voltages for one of the PCBs. 

\subsubsection{Reference sensor response}
\begin{figure}[h]
\bigskip
\centering
\subfigure[]{\includegraphics[width=0.48\textwidth]{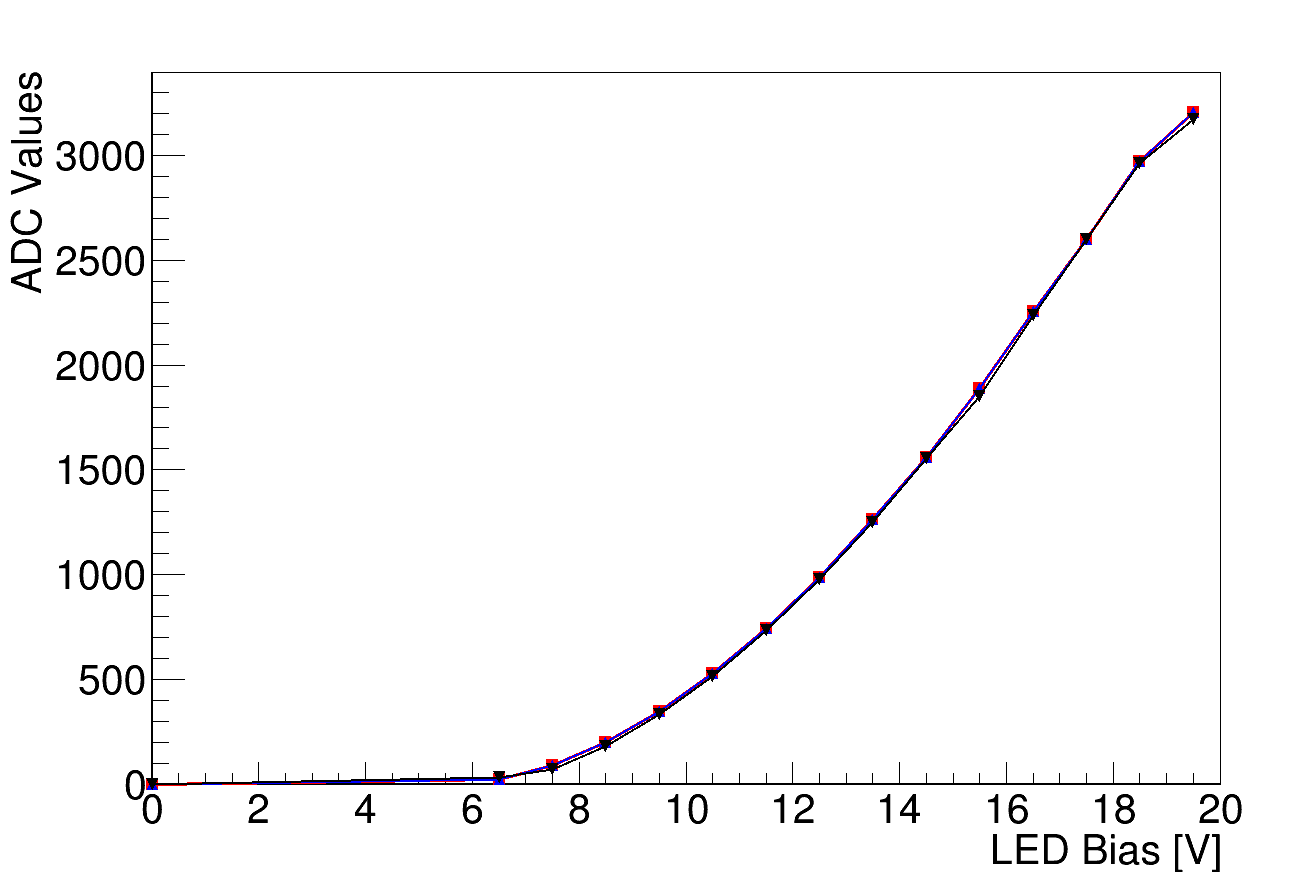}\label{fig:I_SiPMBias}}
\subfigure[]{\includegraphics[width=0.48\textwidth]{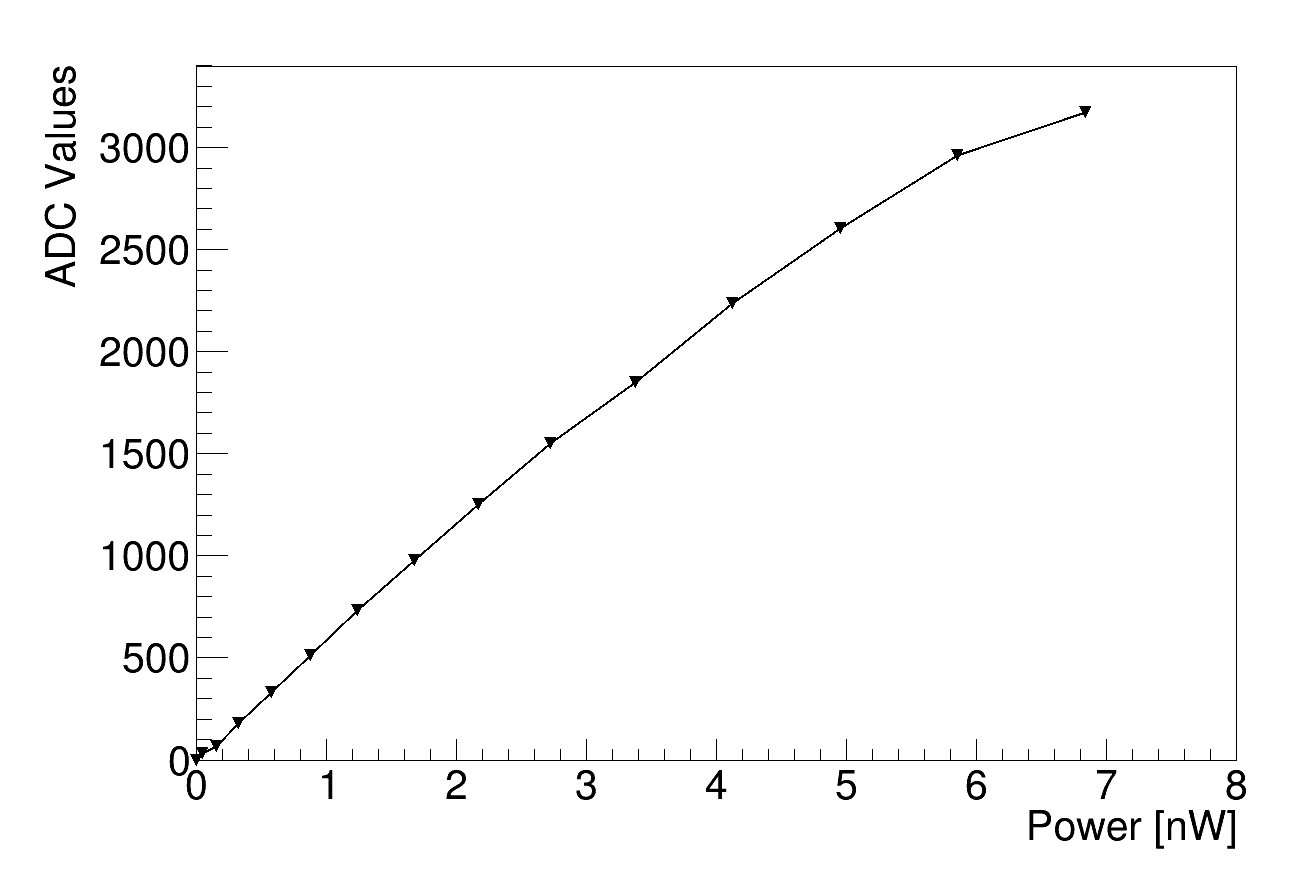}\label{fig:I_PowerADC}}
\caption{(a) Mean ADC value from 1000 events measured with the built-in ADC of the BeagleBone for 3 different repetition rates: 100\,Hz, 1\,kHz and 2\,kHz. It shows the same exponential behaviour as with the powermeter except for very high bias voltages at which saturation of the electronics sets in. The different curves correspond to different pulse repetition rates and no dependency on the frequency is found. (b) The ADC value compared with the simultaneously measured output power at the end of a 1000\,$\mu$m fiber. The error bars are smaller than the markers.}
\label{fig:SiPM}
\end{figure}

The response of the SiPM acting as reference sensor is also studied. The stray light is measured with the SiPM and the built-in ADC of the BeagleBone, while the output power is measured at the end of an optical fiber. The results are shown in Fig.~\ref{fig:SiPM}. The measured signal describes well the output power until $\sim$18 V LED bias voltage, where saturation effects set in.

\subsection{Inner system characterization}
\label{sec3.2}

All the inner components are characterized at RT and CT, using liquid nitrogen (LN$_2$) at 77 K for the cryogenic tests. The expected light transmission from the flange to the PMTs is calculated and compared with the amount of light detected by the PMTs in a dedicated setup.  The light transmission of the fibers is expected to be reduced at CT.

\subsubsection{Fiber and bundle characterization}
\label{sec3.2.1}

 In order to characterize the fibers and bundles individually, they are directly connected to a 460\,nm LED and the light output is measured with a powermeter\footnote{PD300-UV from Ophir} at the end of the fiber.

\begin{figure}[ht]
\bigskip
\centering
 \includegraphics[width=0.48\textwidth]{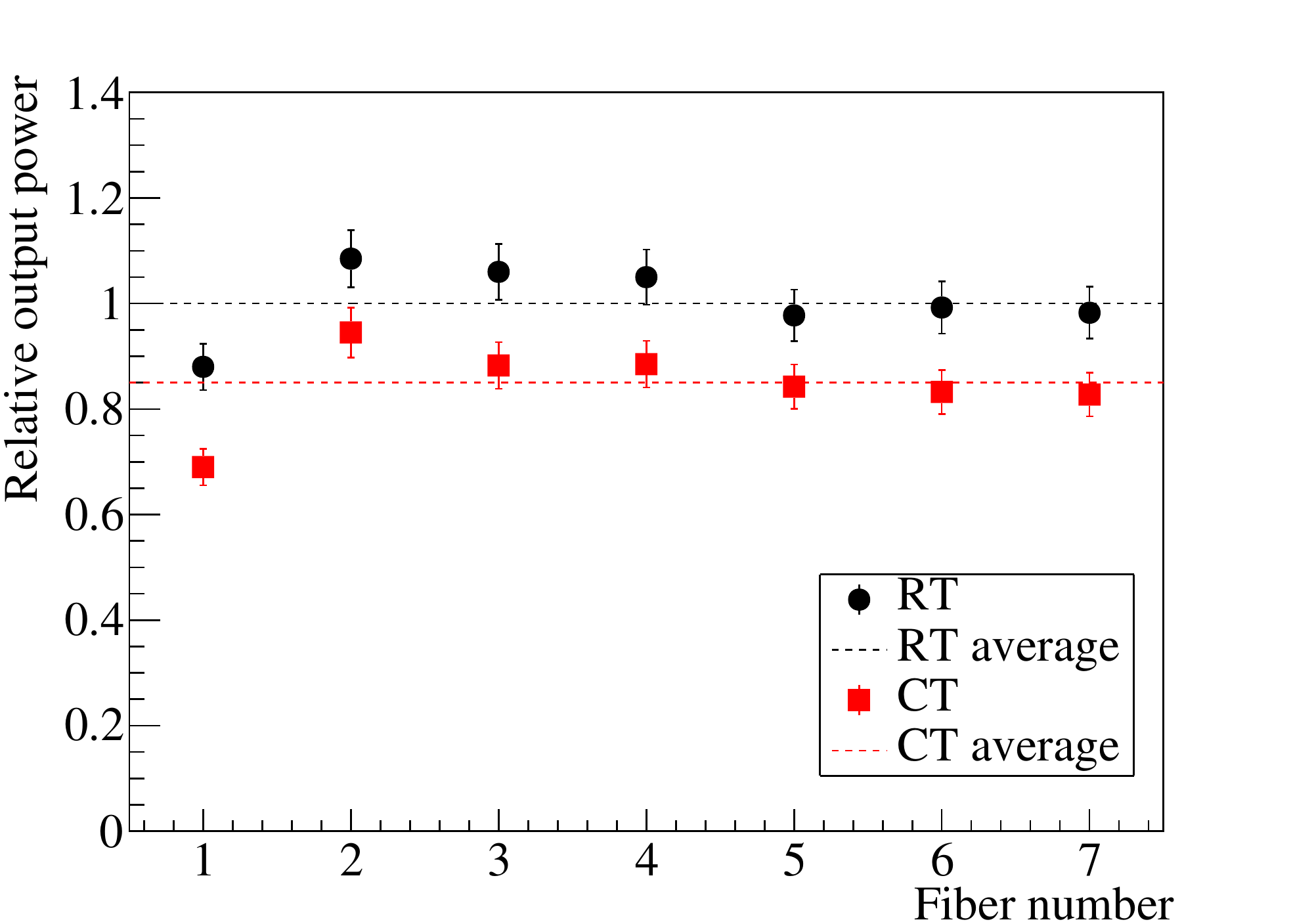}
\caption{Characterization results of the 22.5-m inner fiber at RT and CT. The error bars represent the 5\% systematic error obtained from repetitive measurements. The output power is shown relative to the RT average.}
\label{Fiber}
\end{figure}

The results of the inner fiber characterization are summarized in Fig.~\ref{Fiber}. Considering the dispersion from the average (rms), a homogeneous light output is measured among the fibers, with a light uniformity >90\%. The light attenuation at CT with respect to RT is 0.8\,$\pm$\,0.2\,dB. It is checked that the light transmission at RT is recovered once the LN$_2$ has evaporated. The transmission of the fiber is compared to the one of a shorter fiber. The light attenuation from 1\,m to 22.5\,m is 1.0\,$\pm$\,0.3\,dB at RT, which is consistent with the 1.35\,dB expected from Thorlabs specifications at RT.

\begin{figure}[ht]
\bigskip
\centering
\subfigure[]{\includegraphics[width=0.48\textwidth]{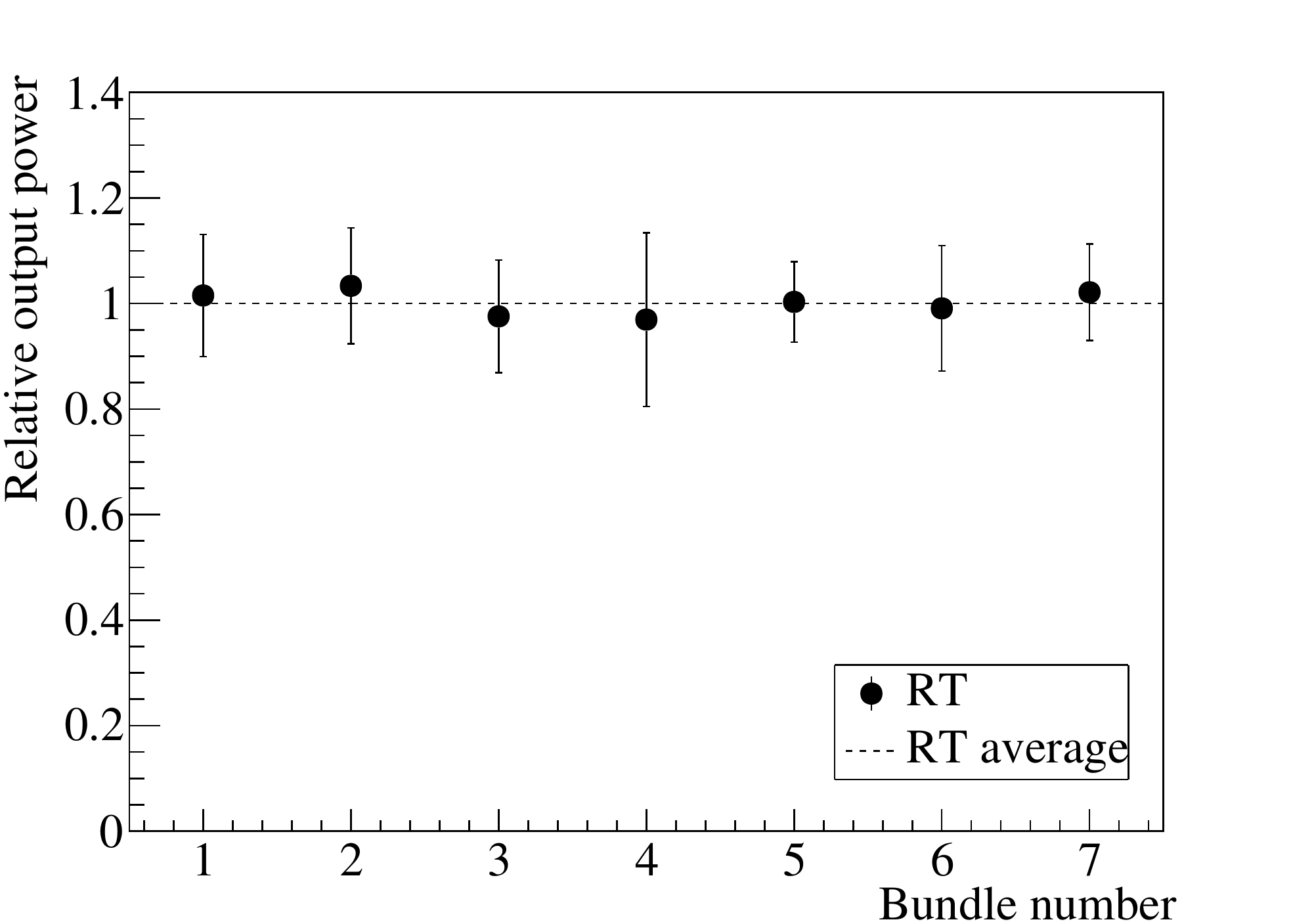}\label{Bundlea}}
\subfigure[]{\includegraphics[width=0.48\textwidth]{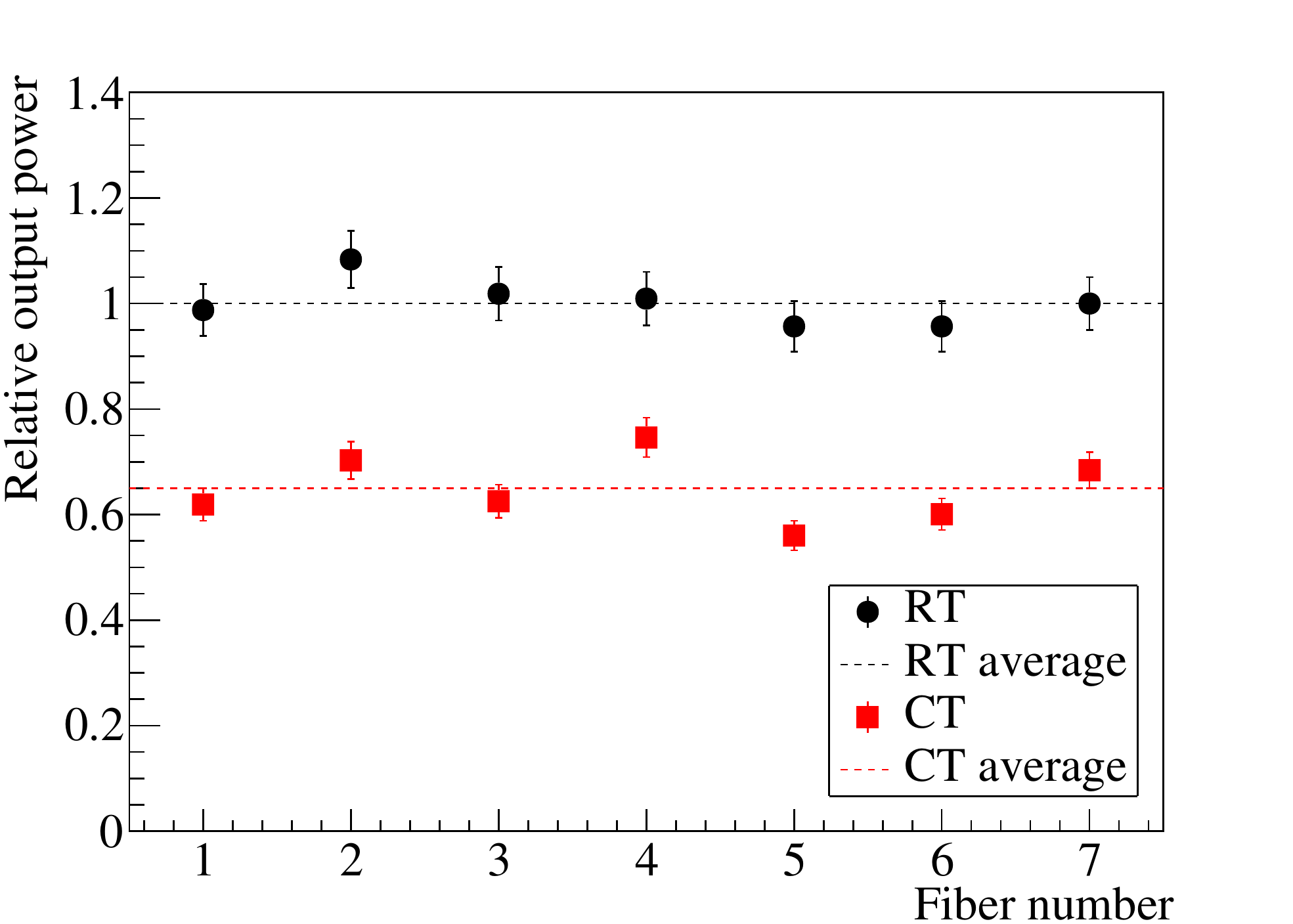}\label{Bundleb}}
\caption{(a) Characterization results of the 1-to-7 fiber bundles at RT. The error bars represent the dispersion among fibers of the same bundle. (b) Bundle characterization at RT and CT, where individual fiber results are given and the error bars represent the 5\% systematic error obtained from repetitive measurements. The output power is shown relative to the RT average.}
\label{Bundle}
\end{figure}

The seven bundles are measured at RT connecting them also between the LED and the powermeter, see Fig.~\ref{Bundlea}. A good homogeneity is also observed among bundles and individual fibers of the bundles, with a light uniformity $\sim$90\%. One bundle is measured at CT to estimate the light loss, see Fig.~\ref{Bundleb}. The light attenuation at CT with respect to RT is 1.9\,$\pm$\,0.3\,dB. As in the case of the fibers, the transmission of the bundle is compared to the performance of a shorter bundle. The bundle transmission variation measured from 3\,m to 1\,m is negligible which is in accordance with the 0.15\,dB expected from Thorlabs.

Finally, the fiber - SMA to SMA vacuum MS - bundle system is measured at RT and CT. To check the effect of the MS at CT, three measurements are carried out: the complete system at RT, everything at CT except the MS, which is outside the dewar at RT, and the complete system at CT. Table~\ref{att} summarizes the attenuation at CT with respect to RT of all the components. In total, 3.1\,$\pm$\,0.3\,dB of attenuation is measured which is consistent with sum of the individual component attenuation (3.1\,$\pm$\,0.4\,dB).

\begin{table}[ht]
\centering
\begin{tabular}{|c|c|}
\hline
Component & A$_{CT}$  \\
			 & (dB)   \\
\hline
\hline
22.5-m fiber & 0.8\,$\pm$\,0.2 \\
3-m bundle 	 & 1.9\,$\pm$\,0.3 \\
MS			 & 0.4\,$\pm$\,0.2 \\
22.5-m fiber + MS + 3-m bundle & 3.1\,$\pm$\,0.3\\
\hline
\end{tabular}
\caption{Measured attenuation by the inner components at CT with respect to operation at RT.}
\label{att}
\end{table}

\subsubsection{Expected light attenuation}
\label{sec3.2.2}

The estimated attenuation of the different inner system components at 465\,nm from the tests performed and the manufacturer specifications are shown in Table~\ref{tab_att}. The fibers and couplings present very few losses. However, the attenuation is dominated by the geometrical efficiency, $\text{A}_\text{Geo}$, which accounts for dividing from one to seven fibers and is estimated with the difference in the surface of the 22.5-m fiber, $\text{S}_{800}$, and each fiber in the bundle, \text{S}$_{200}$:

\begin{equation}
\centering
\text{A}_\text{Geo}= -10\cdot \text{log}\left( \frac{\text{S}_{200}}{\text{S}_{800}} \right) 
= -10\cdot \text{log}\left( \frac{0.031\,\text{mm}^2}{0.503\,\text{mm}^2} \right) = \,12\,\text{dB}
\label{geo}
 \end{equation}
 
 The expected total light attenuation in the detector, including the geometrical inefficiency, is also shown in Table~\ref{tab_att}. The LED must provide a power $\sim$80 times the desired one at the PMT photo-cathode, which is feasible.

\begin{table}[ht]
\centering
\begin{tabular}{|c|c|c|}
\hline
Component   & Attenuation & Reference \\
		      & (dB)  	&  \\
\hline
\hline
MS (flange) 	& <1.5 	& Manufacturer specifications\\
22.5-m fiber 	& 1.0\,$\pm$0.3 	& Sec. \ref{sec3.2.1} \\
MS 				& <1.5 	& Manufacturer specifications\\
Geometry 		& 12	& Equation \ref{geo} \\ 
3-m bundle 		& 0 	& Sec. \ref{sec3.2.1}\\
CT operation	& 3.1\,$\pm$\,0.3 	& Table \ref{att}\\
\hline
\hline
Total at RT & <16.0\,$\pm$\,0.3 &  $\text{A}_\text{MS(Flange)}+\text{A}_\text{22.5m}+\text{A}_\text{MS}+\text{A}_\text{Geo}+\text{A}_\text{3m}$\\
Total at CT & <19.1\,$\pm$\,0.4 &  $\text{A}_\text{MS(Flange)}+\text{A}_\text{22.5m}+\text{A}_\text{MS}+\text{A}_\text{Geo}+\text{A}_\text{3m}+\text{A}_\text{CT}$\\
\hline 
\end{tabular}
\caption{Estimated attenuation of the inner system.}
\label{tab_att}
\end{table}

\subsubsection{Fiber validation with PMTs}
\label{sec3.2.3}

The estimated light attenuation of the total system is compared with the attenuation from dedicated PMT measurements at CT. The setup characterization \cite{protoDUNEPMTs} is used for this test. The measurements are performed simultaneously with 5 R5912-02MOD Hamamatsu PMTs placed inside a 300\,l vessel filled with LN$_2$ at 77\,K. 

The LED is used as an external light source and the amount of light is configurable varying the voltage. The light is guided from the LED to the vessel flange by an optical fiber. The inner system components of the LCS are deployed inside the vessel. The PMT output is measured with a V965A Charge-to-Digital Converter (QDC) from CAEN\footnote{http://www.caen.it/}. The PMTs are biased using a CAEN N1470 power supply. The DAQ is remotely controlled with the aim of automating the data acquisition with LabVIEW software\footnote{4http://www.ni.com/}. 

In order to quantify the light reaching the flange, measurements are performed using a powermeter. The number of photons (ph) at flange is calculated from the measured power as:

\begin{equation}
\# ph = \frac{P\cdot t}{E} = \frac{P/f}{(h\cdot c)/\lambda}
\end{equation}

\noindent where $P$ is the power, $f$ the frequency at which the light source is pulsed, $h$ the Planck constant, $c$ the speed of light, and $\lambda$ the light source wavelength.

\begin{figure}[ht]
\centering
\includegraphics[width=0.48\textwidth]{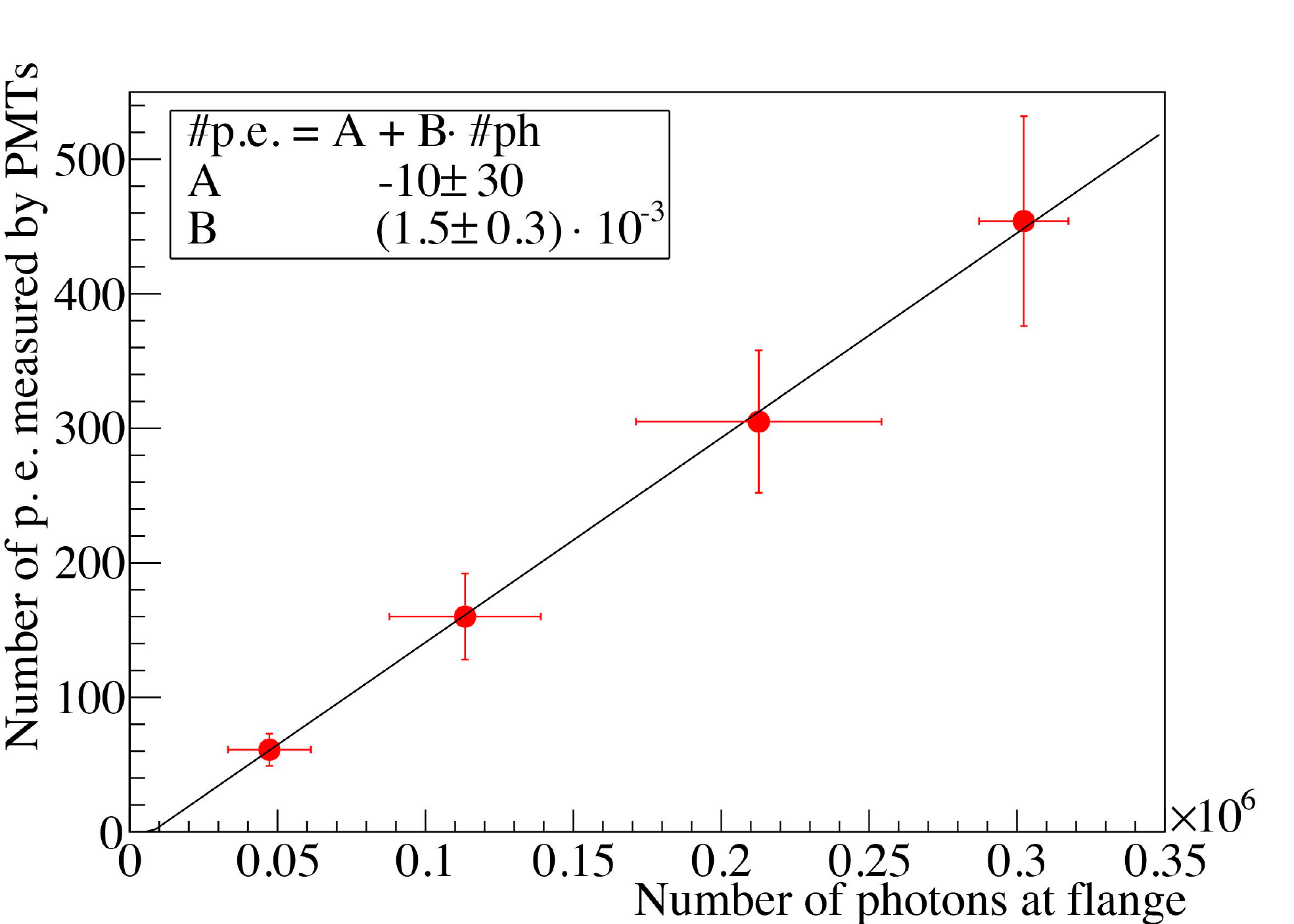}
\caption{Average number of p.e. measured by the PMTs versus the number of photons at the flange. The vertical error bars represent the dispersion among the PMTs tested and the horizontal error bars correspond to the rms of the values measured in two different days.}
\label{plot}
\end{figure}

The amount of incident light is varied from SPE level to hundreds of photoelectrons (p.e.). The plot presented in Fig.~\ref{plot} shows that the attenuation is similar for different light inputs. In Table~\ref{tab_res_pmts}, the light measured by the PMTs over the measured one at flange is compared with the expected attenuation. The attenuation of the inner system in LN$_2$ measured by the PMTs is consistent with the expected one presented in Sec.~\ref{sec3.2.2}.

\begin{table}[ht]
\centering
\begin{tabular}{|c|c|c|}
\hline
\multicolumn{2}{|c|}{Measured by PMTs}&Expected\\
\hline
 \# p.e. / \# ph & Attenuation &Attenuation \\
 (\%) & (dB) & (dB)\\
\hline
\hline
0.15$\pm$0.03	& 20.4\,$\pm$\,0.9 & <19.1\,$\pm$\,0.4\\
\hline
\end{tabular}
\caption{Results of the PMT measurements (from fit in Fig.~\ref{plot}) and expected attenuation at CT (from Sec.~\ref{sec3.2.2}). The attenuation of the inner system components of the LCS is obtained considering 16.5\% PMT quantum efficiency measured at 460\,nm by Hamamatsu for our PMTs at RT, as no difference is expected at CT~\cite{bueno}.}
\label{tab_res_pmts}
\end{table}

\section{Full system validation}
\label{sec4}

Before installation at CERN, the performance of the complete LCS is validated with the final components, and with the inner system at CT. For this test, the same cryogenic system than for the validation shown in Sec.~\ref{sec3.2.3} is used. The goal of the LCS is to measure the PMT gain and to study the PMT response stability. Three different measurements will be performed during the ProtoDUNE-DP operation: gain stability at the operating voltage, gain calibration curve, and PMT response at several photo-electrons level to monitor the quantum efficiency, once the gain is known. The same measurements are carried out during the full system validation. In each case, measurements are taken with one LED providing simultaneously light pulses to 6 PMTs at a time.

The SiPM response is measured with both the QCD and the BeagleBone. The expected correlation between the charge measured by the two devices is observed with <5\% difference, validating the reference sensor.

\begin{figure}[ht]
  \centering
    \centering\includegraphics[width=0.48\textwidth]{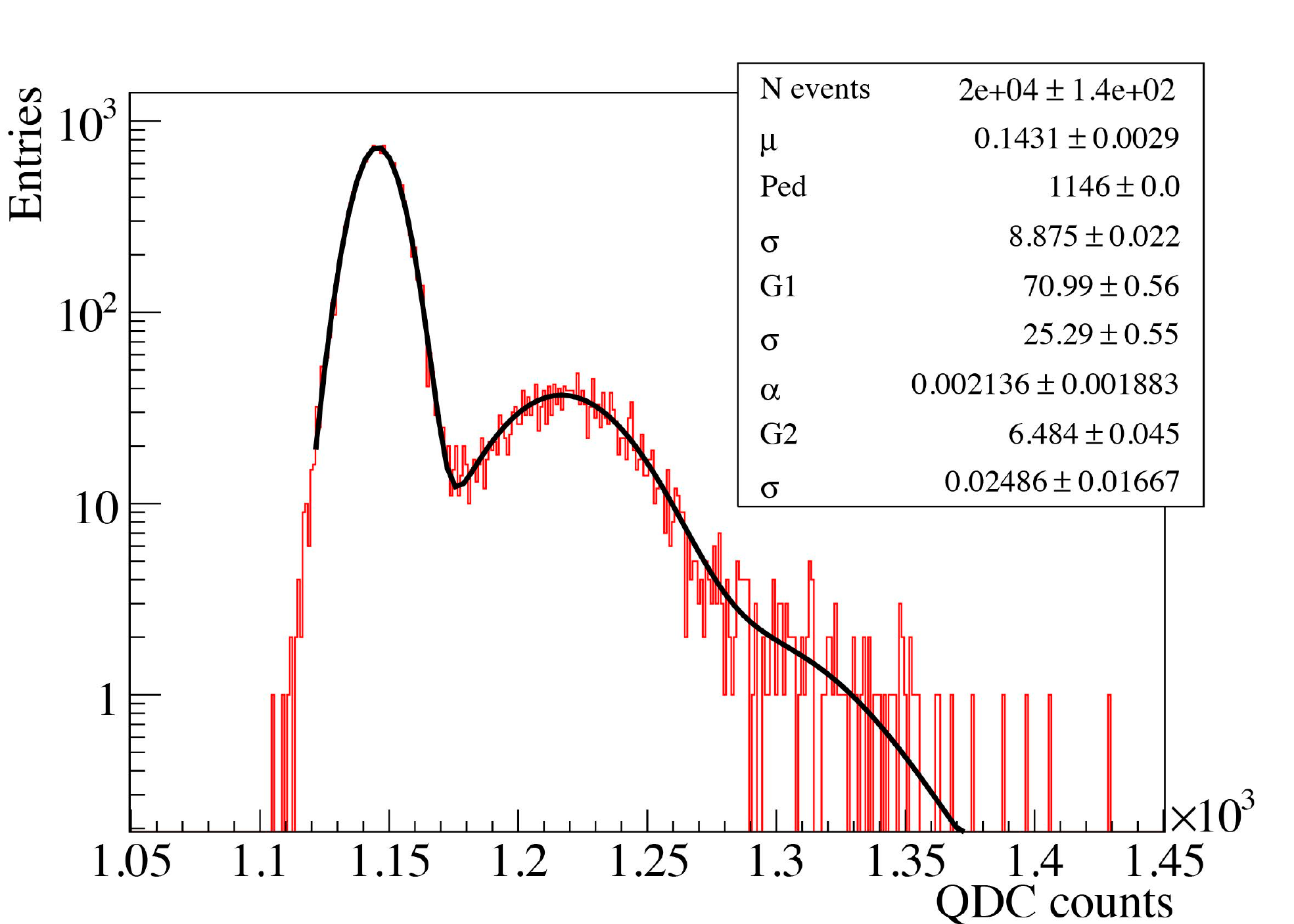}
  \caption{SPE spectrum (red) of a PMT for a $1.2\cdot10^7$ gain taken with the LCS at CT, and fit results (black) where $\mu$ is the mean number of p.e., $Ped$ the position of the pedestal, $G1$ is the amplification of a p.e. from the first dynode (expressed as the mean pulse area), $G2$ from the second dynode, and $\alpha$ is the proportion of p.e. amplified by the second dynode.}
    \label{fits_ct}
\end{figure}

The gain of each PMT is measured taking SPE spectra at the operating HV (corresponding to a gain of $\sim$10$^7$) with the 6 LEDs. An example is shown in Fig.~\ref{fits_ct}. The gain of these PMTs is also measured using a different external LED and a diffuser inside the vessel. Similar gains are obtained and the average discrepancy (7\%) is below the expected uncertainty from the individual PMT characterization (21\%)~\cite{protoDUNEPMTs}. It is concluded that the gain can be well measured using the LCS at CT. Also, the gain curve is mapped successfully with the gain as a function of the high voltage applied to the PMTs.

The response of the PMTs for different light levels can be studied to confirm the optical path integrity, look for relative changes in the PMT quantum efficiency, or as alternative method to determine the PMT gain~\cite{baldini}. On average, the maximum light expected from the LCS at CT is $\sim$60\,p.e. which is enough for the planned measurements. The light output is limited by the maximum LED voltage (19.5\,V) and by the SiPM saturation level. It could be increased if the reference sensor is not needed, i.e. once the correct functioning of the LEDs is ensured. 
 
\section{Outlook for the DUNE DP far detector}
\label{sec5}

The baseline design of the photon detection system for the DUNE DP far detector calls for 720 cryogenic PMTs distributed uniformly on the floor of the cryostat and electrically shielded from the cathode plane~\cite{duneIDRv3} at the bottom of the detector. The 8-inch PMT model (R5912-20Mod from Hamamatsu), the proposed density of PMTs, and their arrangement follows the design of ProtoDUNE-DP detector although detailed modeling and simulations of the light collection performance are ongoing. A LCS will be also required for the DUNE DP far detector PDS. Results from the ProtoDUNE-DP will provide the validation of simulations and will guide optimization for the final detector.

Assuming this design, 120 bundles, 120 fibers, 120 light sources, 120 feedthroughs, and 20 reference sensors will be needed. The length of the fibers and bundles has to be calculated considering the exact position of the feedthrough flanges. However, alternatives to this design will be pursued with R\&D measurements in order to minimize the amount of fibers, study other options for the reference sensor, and increase the input light if necessary. The number of fibers, and therefore the cost of the system, can be reduced by using light dffusers, so that one fiber can illuminate at least 4 PMTs. Another option to reduce the number of LEDs is to install a 1-to-7 fiber bundle directly from the LED to each feedthrough. This test was carried out during the full system validation and the conclusion is that it would be possible to calibrate the 36 PMTs using only one LED as enough light would reach the PMTs. This will be considered as a possible option for the DUNE DP far detector and will be tested in ProtoDUNE-DP.
\section*{Conclusions}
\addcontentsline{toc}{section}{\protect\numberline{}Conclusions}%
\label{sec6}

The ProtoDUNE-DP experiment aims to build and operate a 300\,t LAr TPC at CERN to fully demonstrate the dual-phase technology at large scale for DUNE, the next generation long-baseline neutrino experiment. The photon detection system will add timing capabilities, and will be formed by 8-inch cryogenic PMTs from Hamamatsu positioned at the bottom of the detector. An LED-based fiber calibration system has been designed and validated for the ProtoDUNE-DP PMTs. The goal of the LCS is to measure the PMT gain and study the PMT response stability. The calibration light is provided by a blue LED of 465\,nm using a Kapustinsky circuit as LED driver and a fiber system ends with an optical fiber installed at each PMT. There is also a reference sensor to check the LED performance. Each LED is connected to an external fiber going to one feedthrough, inner fibers are connected inside the cryostat and each one of these fibers is attached to a 1-to-7 fiber bundle, so that one fiber is finally installed pointing at each PMT. Characterization measurements have validated the external and internal components individually, and light losses due to couplings, fiber length or cryogenic operations are minimal. A prototype reproducing the full-scale system has been tested with PMTs and the PMT gain calibration has been performed successfully. The LCS is ready to be installed and operated in the ProtoDUNE-DP detector at CERN.

\acknowledgments

This project has received funding from the European Union Horizon~2020 Research and Innovation programme under Grant Agreement no.~654168 and from the Spanish Ministerio de Economia y Competitividad (SEIDI-MINECO) under Grants no.~FPA2016-77347-C2-1-P, FPA2016-77347-C2-2-P, MdM-2015-0509, and SEV-2016-0588.



\bibliographystyle{JHEP}
\bibliography{main}



\end{document}